# Ablation of solids by femtosecond lasers: ablation mechanism and ablation thresholds for metals and dielectrics

E. G. Gamaly*[1] A. V. Rode[1], V. T. Tikhonchuk[2], and B. Luther-Davies[1]

[1]*Research School of Physical Science and Engineering,*

*Australian National University, Canberra, ACT 0200 Australia*

[2]*P.N. Lebedev Physical Institute, Moscow, Russia*

## ABSTRACT

The mechanism of ablation of solids by intense femtosecond laser pulses is described in an explicit analytical form. It is shown that at high intensities when the ionization of the target material is complete before the end of the pulse, the ablation mechanism is the same for both metals and dielectrics. The physics of this new ablation regime involves ion acceleration in the electrostatic field caused by charge separation created by energetic electrons escaping from the target. The formulae for ablation thresholds and ablation rates for metals and dielectrics, combining the laser and target parameters, are derived and compared to experimental data. The calculated dependence of the ablation thresholds on the pulse duration is in agreement with the experimental data in a femtosecond range, and it is linked to the dependence for nanosecond pulses.

[*] e-mail: gam110@rsphysse.anu.edu.au; ph.: ++61-2-6125-0171; fax: ++61-2-6125-0732

I. INTRODUCTION:

THE ULTRA SHORT PULSE LASER-MATTER INTERACTION MODE

The rapid development of femtosecond lasers over the last decade has opened up a wide range of new applications in industry, material science, and medicine. One important physical effect is material removal or laser ablation by femtosecond pulses which can be used for the deposition of thin films; the creation of new materials; for micro-machining; and, in the arts, for picture restoration and cleaning. Femtosecond laser ablation has the important advantage in such applications compared with ablation using nanosecond pulses because there is little or no collateral damage due to shock waves and heat conduction produced in the material being processed. In order to choose the optimal laser and target parameter it is useful to have simple scaling relations, which predict the ablation condition for an arbitrary material. In this paper we present an analytical description of the ablation mechanism and derive appropriate analytical formulae.

In order to remove an atom from a solid by the means of a laser pulse one should deliver energy in excess of the binding energy of that atom. Thus, to ablate the same amount of material with a short pulse one should apply a larger laser intensity approximately in inverse proportion to the pulse duration. For example, laser ablation with 100 fs pulses requires an intensity in a range $\sim 10^{13} - 10^{14}$ W/cm$^2$ [1], while 30-100 ns pulse ablates the same material at the intensities $\sim 10^8 - 10^9$ W/cm$^2$ [2]. At intensities above $10^{13} - 10^{14}$ W/cm$^2$ ionization of practically any target material takes place early in the laser pulse time. For example, if an intense, $10^{13} - 10^{14}$ W/cm$^2$, femtosecond pulse interacts with a dielectric, almost full single ionization of the target occurs at the beginning of the laser pulse. Following ionization, the laser energy is absorbed by free electrons due to inverse Brehmstrahlung and resonance absorption mechanisms and does not depend on the initial state of the target. Consequently, the interaction with both metals and dielectrics proceeds in a similar way which contrasts to the situation when a long pulse is where ablation of metals occurs at relatively low intensity compared with that for a transparent dielectric whose absorption is negligibly small.

Another distinctive feature of the ultra short interaction mode is that the energy transfer time from the electrons to ions by Coulomb collisions is significantly longer (picoseconds) that the laser pulse duration ($t_p \sim 100$ fs). Therefore, the conventional hydrodynamics motion does not occur during the femtosecond interaction time.

There are two forces which are responsible for momentum transfer from the laser field and the energetic electrons to the ions in the absorption zone: one is due to the electric field of charge separation and another is the ponderomotive force. The charge separation occurs if the energy absorbed by the electrons exceeds the Fermi energy, which is approximately a sum of the binding energy and work function, so the electrons can escape from the target. The electric field of charge separation pulls the ions out of the target. At the same time, the ponderomotive force of the laser field in the skin layer pushes electrons deeper into the target. Correspondingly it creates a mechanism for ion acceleration into the target. Below we demonstrate that the former mechanism dominates the ablation process for sub-picosecond laser pulses at an intensity of $10^{13} - 10^{14}$ W/cm$^2$. This mechanism of material ablation by femtosecond laser pulses is quite different from the thermal ablation by long pulses.

Femtosecond ablation is also sensitive to the temporal and spatial dependence of the intensity of the laser pulse. The Chirped Pulse Amplification (CPA) technique commonly used for short pulse generation [3] can produce a main (short) pulse accompanied by a nanosecond pre-pulse or pedestal that can be intense enough itself to ablate the target. Therefore, an important condition for the practical realization of the pure femtosecond interaction mode should be that the intensity in any pre-pulse has to be lower than the thresholds for ablation or ionization in the nanosecond regime. There are fortunately several methods for achieving high pulse contrast (nonlinear absorbers, conversion to second harmonic, etc.) [4,5].

A rather simple and straightforward analytical model can describe the ultra short pulse mode of laser-matter interaction. The main features of this model were developed more than 10 years ago in connection with the ultra short and super intense laser-matter interaction [5]. In what follows this model is modified and applied to the problems of the laser ablation at relatively moderate intensities near the ablation threshold for solids. The absorption coefficient, ionization and ablation rates, and ablation thresholds for both metals and dielectrics are expressed in terms of the laser and target parameters by explicit formulae. The comparison to the long-pulse interaction mode and to the experimental data is presented and discussed.

## 2. LASER FIELD PENETRATION INTO A TARGET: SKIN-EFFECT

The femtosecond laser pulse interacts with a solid target with a density remaining constant during the laser pulse (the density profile remains step-like). The laser electromagnetic field in the target (metal or plasma) can be found as a solution to Maxwell equations coupled to the material equations. The cases considered below fall in framework of the normal skin-effect [5,6] where the laser electric field $E(x)$ decays exponentially with the depth into the target:

$$E(x) = E(0)\exp\left[-\frac{x}{l_s}\right]; \quad (1)$$

here $l_s$ is the field penetration (or absorption) length (skin-depth); the target surface corresponds to $x = 0$, and Eq.(1) is valid for $x > 0$. The absorption length in general is expressed as follows [6]:

$$l_s = \frac{c}{\omega k} \quad (2)$$

where $k$ is the imaginary part of the refractive index, $N = \varepsilon^{1/2} = n + ik$ ($\varepsilon = \varepsilon' + i\varepsilon''$ is the dielectric function and $\omega$ is the laser frequency). We take the dielectric function in the Drude form for the further calculations:

$$\varepsilon = 1 - \frac{\omega_{pe}^2}{\omega(\omega + i\nu_{eff})} \quad (3)$$

where $\omega_{pe} = (4\pi e^2 n_e / m_e)^{1/2}$ is the electron plasma frequency, and $\nu_{eff}$ is an effective collision frequency of electrons with a lattice (ions). In the case of a high collision rate $\nu_{eff} \gg \omega$ and thus $\varepsilon'' \gg \varepsilon'$ one can reduce Eq.(2) to the conventional skin depth expression for the high-conducting metals [6]:

$$l_s = \frac{c}{\omega k} \approx \frac{c}{\omega_{pe}} \left( \frac{2\nu_{eff}}{\omega} \right)^{1/2}. \qquad (4)$$

The main difference between these formulae and those ones for the conventional low-intensity case resides in the fact that the real and imaginary parts of the dielectric permittivity, and thus, the plasma frequency and the effective collision frequency, all are intensity and time-dependent. The finding of these dependencies is the subject of next sections.

## 3. ABSORPTION MECHANISMS AND ABSORPTION COEFFICIENT

The light absorption mechanisms in solids are the following [7]:

1. intraband absorption, and contribution of free charge carriers in metals and semiconductors;
2. interband transitions and molecular excitations;
3. absorption by collective excitations (excitons, phonons);
4. absorption due to the impurities and defects.

At high intensities $\sim 10^{13} - 10^{14}$ W/cm$^2$, the electron oscillation energy in the laser electric field is a few electron-volts, which is of the order of magnitude of the ionization potential. Futhermore, at intensities above $10^{14}$ W/cm$^2$ the ionization time for a dielectric is just a few femtoseconds, typically much shorter than the pulse duration (~100 fs). The electrons produced by ionization then in dielectrics dominate the absorption in the same way as the free carriers in metals, and the characteristics of the laser-matter interaction become independent of the initial state of the target. As a result the first mechanism becomes of a major importance for both metals and dielectrics. In the presence of free electrons, inverse Bremsstrahlung and resonant absorption (for *p*-polarized light at oblique incidence) become the dominant absorption mechanisms.

However, one should not oversimplify the picture. The electron interaction with the lattice (the electron-phonon interaction) and the change in the electron effective mass might be significant in dielectrics and even in some metals [11]. The number density of conductivity electrons in metal changes during the pulse. We should also note that in many cases the real part of the dielectric function is comparable to the imaginary part. Then the skin-effect solution (for example, the simple formula Eq.4) should be replaced by a more rigorous approach. We use below the Fresnel formulae [8] with the Drude-like dielectric function for the absorption coefficient calculations taking into account that the density of the target during the pulse remains unchanged. Then the conventional formulae for the reflection $R$ and absorption $A$ coefficients are the following [8]:

$$R = \frac{(n-1)^2 + k^2}{(n+1)^2 + k^2}; \quad A = 1 - R. \tag{5}$$

In the limit of low absorption $A \ll 1$, which holds for the high conductivity perfect metals, one finds a simple expression the absorption coefficient (cf. Appendix A [3]):

$$A \approx \frac{2\omega l_s}{c}. \tag{6}$$

It should be noted that the skin-depth is the function of the laser intensity and time. This function we shall find below.

## 4. INTENSITY THRESHOLD FOR IONISATION OF DIELECTRICS

The dielectric function in dielectrics at low intensity of the electric field is characterized by the large real and small imaginary parts. The imaginary part increases mainly due to ionization. There are two major mechanisms of ionization in the laser field: ionization by electron impact (avalanche ionization); and the multiphoton ionization. The time dependence of the number density of free electrons $n_e$ stripped off the atoms by these processes is defined by the rate equation [1,9]:

$$\frac{dn_e}{dt} = n_e w_{imp} + n_a w_{mpi} \tag{7}$$

here $n_a$ is the density of neutral atoms, $w_{imp}$ is the time independent probability (in s$^{-1}$) for the ionization by electron impact, and $w_{mpi}$ is the probability for the multiphoton ionization [5,9]. For the case of single ionization it is convenient to present the probabilities $w_{imp}$ and $w_{mpi}$, in the form:

$$w_{imp} \approx \frac{\varepsilon_{osc}}{J_i}\left(\frac{2\omega^2 v_{eff}}{\omega^2 + v_{eff}^2}\right); \tag{8}$$

$$w_{mpi} \approx \omega n_{ph}^{3/2}\left(\frac{\varepsilon_{osc}}{2J_i}\right)^{n_{ph}}; \tag{9}$$

here, $\varepsilon_{osc}$ is the electron quiver energy in the laser field, $n_{ph} = \hbar\omega/J_i$ is the number of photons necessary for atom ionization by the multiphoton process, $J_i$, is the ionization potential, and $v_{eff}$ is the effective collision frequency. It must be emphasised that the effective collision frequency in Eq. (8) accounts for the inelastic collisions leading to the energy gain by the electrons. In general, it differs from the effective collision frequency in Eq. (3) and in the Sections below, which accounts for the momentum exchange due to elastic collisions.

One can see from the Eqs. (8) and (9) that the relative role of the impact and multiphoton ionization depends dramatically on the relation between the electron quiver energy and the ionization potential. If $\varepsilon_{osc} > J_i$ then $w_{mpi} > w_{imp}$, and the multiphoton ionization dominates for any relationship between the frequency of the incident light and the efficient collision frequency. By presenting the oscillation energy in a scaling form:

$$\varepsilon_{osc}[\text{eV}] = 9.3\left(1+\alpha^2\right)\frac{I}{10^{14}[\text{W/cm}^2]}\left(\lambda[\mu m]\right)^2 \tag{10}$$

where $\alpha$ accounts for the laser polarization ($\alpha = 1$ for the circular and $\alpha = 1$ for the linear polarisation), it is evident that the multiphoton ionization dominates in the laser-interaction at intensities above $10^{14}$ W/cm$^2$ (for the 100-fs pulse duration this condition corresponds to the laser fluence of 10 J/cm$^2$.)

The general solution to Eq. (7) with the initial condition $n_e(t = 0) = n_0$ is the following:

$$n_e(I,\lambda,t) = \left\{ n_0 + \frac{n_a w_{mpi}}{w_{imp}} \left[ 1 - \exp(-w_{imp}t) \right] \right\} \exp(w_{imp}t) \qquad (11)$$

It is in a good agreement with the direct numerical solution to the full set of kinetic equations [1]. Electron impact ionization is the main ionization mechanism in the long (nanosecond) pulse regime. Then $\varepsilon_{osc} \ll J_i$ and $\omega \ll v_{eff}$, and one can neglect the second term in Eq. (11) and the number of free electrons exponentially increases with the product of $w_{imp}$ and the pulse duration: $n_e \sim n_0 \exp\{w_{imp} \times t_p\}$. Therefore in the long-pulse regime the ionization threshold depends on the laser fluence $F = I \times t_p$. In the case of high intensities multiphoton ionization dominates and the number of free electrons increases linearly with time, $n_e \sim n_a w_{mpi} t$. In this case the ionization time could be shorter than the pulse duration and the ionization threshold depends on the laser intensity and laser wavelength. It is conventional to suggest that the ionization threshold (or breakdown threshold) be achieved when the electron number density reaches the critical density corresponding to the incident laser wavelength [7]. The ionization threshold for the majority of materials lies at intensities in between ($10^{13}$ - $10^{14}$) W/cm$^2$ ($\lambda \sim 1$ µm) with a strong nonlinear dependence on intensity. For example, for a silica target at the intensity $2 \times 10^{13}$ W/cm$^2$ avalanche ionisation dominates, and the first ionisation energy is not reached by the end of 100-fs pulse at 1064 nm. At $10^{14}$ W/cm$^2$ multiphoton ionisation dominates and the full ionisation is completed after 20 fs. It should be also mentioned that the ionisation threshold decreases with the increase in the photon energy.

## 5. ELECTRON COLLISION FREQUENCY AND ENERGY TRANSFER FROM ELECTRONS TO IONS

When the ionisation is completed, the plasma formed in the skin-layer of the target has a free-electron density comparable to the ion density of about $10^{23}$ cm$^{-3}$. In order to meet the ablation conditions the average electron energy should increase up to the Fermi energy $\varepsilon_F$, i.e. up to several eV. The electron-electron equilibration time is of the order of magnitude of the reciprocal electron plasma frequency, i.e. $\sim \omega_{pe}^{-1} \sim 10^{-2}$ fs that is much shorter than the pulse duration. Therefore the electron energy distribution is close to equilibrium and follows the laser intensity evolution in time adiabatically adjusting to any changes. The electron gas is non-ideal in the high-density conditions: the energy of Coulomb interaction is comparable to the electron kinetic energy and there are only few electrons in the Debye sphere. There are no reliable analytical expressions for the effective electron collision frequency in this energy-density domain. There are interpolation formulae for some materials in [10]. Physically sound estimates could also be made.

The effective electron-ion collision frequency could be estimated by approaching the Fermi energy from two limiting cases: from the low and from the high temperature limits. In the low-temperature limit the electron-phonon collisions dominate. The electron-phonon collision frequency increases in direct proportion with the temperature for $T$ above the Debye temperature $T_D$. In the opposite high-temperature limit the effective frequency of electron-ion Coulomb collisions decreases with the electron temperature. Thus, the effective collision frequency has maximum at the electron temperature approaching Fermi energy. The electron-phonon collision frequency (or, the probability for an electron to emit or to absorb a phonon) one can estimate at the temperature $T_D \ll T \ll T_F$ as the following [7,10]:

$$\nu_{e-ph} \approx \left(\frac{m_e}{M}\right)^{1/2} \frac{J_i}{\hbar} \frac{T}{T_D} \qquad (12)$$

Taking, for example $J_i = 7.7$ eV (first ionization potential for copper), $T_D$(Cu) ~ 300 K, and $M_{Cu}$

= 63.54 a.m.u. one obtains $v_{eff} \sim 9\times10^{15}$ s$^{-1}$. This estimate is very close to the effective frequency at the room temperature $8.6\times10^{15}$ s$^{-1}$ extracted from the conductivity measurement [11].

At high temperatures $T_e >> \varepsilon_F$ the effective electron-ion collision frequency could be estimated by using the model for an ideal plasma at solid state density. The collision is considered as a 90-degree deflection of an electron path due to the Coulomb interaction with the ion, and the collision frequency is the frequency of the momentum exchange. According to [10]:

$$v_{ei} \approx 3\times10^{-6} \ln\Lambda \frac{n_e Z}{T_{eV}^{3/2}}. \tag{13}$$

For example, from Eq.13 the electron-ion collision frequency in Copper at the electron temperature coinciding with the Fermi energy ($n_e = 0.845\times10^{23}$ cm$^{-3}$, $\omega_p = 1.64\times10^{16}$ s$^{-1}$, $T_{eV} \sim 7.7$ eV, $\ln\Lambda \sim 2$) is $v_{ei} = 2.38\times10^{16}$ s$^{-1}$. This value is about twice higher than that estimated from the low temperature case, and almost coincides with the plasma frequency, $v_{eff} \sim \omega_{pe} = 2.39\times10^{16}$ s$^{-1}$. It seems reasonable to assume that $v_{eff} \approx \omega_{pe}$ for the further estimates of the ablation threshold, as it has been suggested in [1]. The value of $v_{eff}$ can be corrected by experimental measurements of the skin depth (ablation depth). Some more advanced models and interpolations for the effective collision frequency were derived in [10].

Thus, in the ablation conditions $v_{ei} >> \omega$. Therefore the electron mean free path is much smaller than the skin depth. That is, the condition for the normal skin effect is valid.

The electron-ion energy transfer time in a dense plasma can be expressed through the collision frequency (13) as follows:

$$\tau_{ei} \approx \frac{M}{m_e} v_{ei}^{-1} \tag{14}$$

The estimation for copper yields the ion heating time $\tau_{ei} = 4.6\times10^{-12}$ s, which is in agreement with the values suggested by many authors [1,5,13]. A similar estimate for silica gives $6.4\times10^{-12}$ s. Therefore, for the sub-picosecond pulses ($t_p \sim 100$ fs) the ions remain cold during the laser pulse interaction with both metals and dielectrics.

## 6. ELECTRON HEATING IN THE SKIN LAYER

In the previous Section we have demonstrated that electrons have no time to transfer the energy to the ions during the laser pulse $\tau_{ei} > t_p$. That means that the target density remains unchanged during the laser pulse. The electrons also cannot transport the energy out of the skin layer because the heat conductivity time is much longer than the pulse duration. It is easy to see that the electron heat conduction time $t_{heat}$ (the time for the electron temperature smoothing across the skin-layer $l_s$) is also much longer than the pulse duration. Indeed, the estimates for this time with the help of conventional thermal diffusion [6] give:

$$t_{heat} \approx \frac{l_s^2}{\kappa}; \quad \kappa = \frac{l_e v_e}{3};$$

here $\kappa$ is coefficient for thermal diffusion, $l_e$ and $v_e$ are the electron mean free path and velocity correspondingly. Using Copper as an example yields $l_s = 67.4$ nm, $\kappa \sim 1$ cm$^2$/s, and the electron heat conduction time is in the order of tens of picoseconds.

The energy conservation law takes a simple form of the equation for the change in the electron energy $T_e$ due to absorption in a skin layer [5]:

$$c_e(T_e) n_e \frac{\partial T_e}{\partial t} = -\frac{\partial Q}{\partial x}; \quad Q = A I_0 \exp\left\{-\frac{2x}{l_s}\right\}; \tag{15}$$

here $Q$ is the absorbed energy flux in the skin layer, $A = I/I_0$ is the absorption coefficient, $I_0 = cE^2/4\pi$ is the incident laser intensity, $n_e$ and $c_e$ are the number density and the specific heat of the conductivity electrons. In a simple model of the ideal Fermi gas the electron specific heat increases with electron temperature from the low-temperature level $c_e = \pi^2 T_e/2\varepsilon_F$ for $T_e \ll \varepsilon_F$ [11] up to the maximum value of $c_e \sim 3/2$ for the conventional ideal gas at high temperature $T_e > \varepsilon_F$. The specific heat could also be found as a tabulated function corrected on the experimental measurements, which are usually evidencing the deviations from the simple model of the ideal

Fermi gas [11]. The absorption coefficient and the skin depth are the known functions of the incident laser frequency $\omega$, the number density of the conductivity electrons $n_e$, (or, plasma frequency $\omega_{pe}$), the effective collision frequency including electron-ion and electron-phonon collisions $v_{eff}$, the angle of the incidence, and polarisation of the laser beam [5]. In fact, both material parameters $\omega_{pe}$ and $v_{eff}$, are temperature dependent. Therefore, Eq. (15) describes the skin effect interaction with the time-dependent target parameters. In order to obtain convenient scaling relations for the ablation rate we use, as a first approximation, the conventional skin effect approach with time-independent characteristics and with the specific heat of the ideal gas. Such an approximation is applicable because at the ablation threshold $T_e \approx \varepsilon_F$. Thus, the time integration of the Eq.(15) yields time and space dependencies of the electron energy in the skin layer:

$$T_e = \frac{4}{3} \frac{A I_0 t}{l_s n_e} \exp\left\{-\frac{2x}{l_s}\right\}; \quad T_e \approx \varepsilon_F. \qquad (16)$$

This approach is well justified for metals because the temperature dependent skin-depth and absorption coefficient enter into the above formula as a ratio $A/l_s$, which is a weak function of temperature. Indeed, in the low-absorption case (A<<1) for the highly conductive perfect metals, the absorption coefficient expresses by (6), and the ratio $A/l_s$ is almost constant:

$$\frac{A}{l_s} \approx \frac{2\omega}{c}. \qquad (17)$$

In the high absorption case this ratio changes weakly being of the same order of magnitude with the correction factor of ~ 1.3 (see Appendix A). The number density of the conductivity electrons is also almost constant during the interaction time.

The relationship Eq.(16) represents an appropriate scaling law for the electron temperature in the skin layer. The experimental data correlate well with the prediction of the Eq.(16). For example, the estimate of the skin depth in a Copper target irradiated by a Ti:sapphire laser ($\lambda$ = 780 nm, $\omega$ = 2.4×10$^{15}$ s$^{-1}$; $v_{eff} \approx \omega_{pe}$ = 1.639×10$^{16}$ s$^{-1}$, $n_e$ = 0.845×10$^{23}$

cm$^{-3}$) gives with the help of Eq.(4) $l_s$ = 67.4 nm. The maximum electron temperature at the surface of the Copper target under the fluence $AI_0 t_p$ = 1 J/cm$^2$ reaches $T_e$ = 7,5 eV, which is close to the Fermi energy for Copper.

## 7. ABLATION MECHANISM:
## IONS PULLED OUT OF THE TARGET BY ENERGETIC ELECTRONS

It has been shown in the preceding section that the free electrons in the skin layer can gain the energy exceeding the threshold energy required to leave a solid target during the pulse time. The energetic electrons escape the solid and create a strong electric field due to charge separation with the parent ions. The magnitude of this field depends directly on the electron kinetic energy $\varepsilon_e \sim (T_e - \varepsilon_{esc})$ ($\varepsilon_{esc}$ is the work function) and on the gradient of the electron density along the normal to the target surface (assuming one dimensional expansion) [6,14]:

$$E_a = -\frac{\varepsilon_e(t)}{e}\frac{\partial \ln n_e}{\partial z}. \qquad (18)$$

A ponderomotive force of the electric field in the target is another force applied to the ions during the laser pulse [15]:

$$F_{pf} = -\frac{2\pi e^2}{m_e c \omega^2}\nabla I.$$

However, for the solid density plasma and at the intensities of ~10$^{14}$ W/cm$^2$ it is significantly smaller than the electrostatic force $eE_a$.

The field Eq.(18) pulls the ions out of the solid target if the electron energy is larger than the binding energy, $\varepsilon_b$, of ions in the lattice. The maximum energy of ions dragged from the target reaches: $\varepsilon_i(t) \approx Z\varepsilon_e(t) \approx (T_e - \varepsilon_{esc} - \varepsilon_b)$. The time necessary to accelerate and ablate ions could be estimated with the help of the equation for the change of ion momentum:

$$\frac{dp_i}{dt} \approx eE_a. \qquad (19)$$

The characteristic scale length for the expanding electron cloud is the Debye length $l_D \sim v_e/\omega_{pe}$, where $v_e = [(T_e - \varepsilon_{esc})/m_e]^{1/2}$ is the electron thermal velocity. Thus, the ion acceleration time, i.e. the time required for the ion to acquire the energy of $\varepsilon_e$ could be found with the help of (19) as the following:

$$t_{acc} = \frac{l_D}{v_i} \approx \frac{2}{\omega_{pe}} \left(\frac{m_i}{m_e}\right)^{\frac{1}{2}} \left(\frac{T_e - \varepsilon_{esc}}{T_e - \varepsilon_{esc} - \varepsilon_b}\right)^{\frac{1}{2}}. \qquad (20)$$

Below the ablation threshold, when $T_e \sim \varepsilon_{esc} + \varepsilon_b$ the acceleration time is much longer than the pulse duration. However, when the laser fluence exceeds the ablation threshold this time is comparable and even shorter than the pulse duration. For example, for Copper at $F = 1$ J/cm² this time is less than 40 fs. This means that for high intensities (fluences) well over the ablation threshold the equation (15) for the electron temperature should include the energy losses for ion heating. This effect of electrostatic acceleration of ions is well known from the studies of the plasma expansion [6,14] and ultrashort intense laser-matter interaction [5].

## A. Ablation threshold for metals

According to Eq. (20), the minimum energy that electron needs to escape the solid equals to the work function. In order to drag ion out of the target the electron must have an additional energy equal to or larger than the ion binding energy. Hence, the ablation threshold for metals can be defined as the following condition: the electron energy must reach, in a surface layer $d << l_s$ by the end of the laser pulse, the value equal to the sum of the atomic binding energy and the work function. Using the Eq.(16) for the electron temperature we obtain the energy condition for the ablation threshold:

$$\varepsilon_e = \varepsilon_b + \varepsilon_{esc} = \frac{4}{3}\frac{AI_0 t_p}{l_s n_e}. \qquad (21)$$

The threshold laser fluence for ablation of metals is then defined as the following:

$$F_{th}^m \equiv I_0 t_p = \frac{3}{4}(\varepsilon_b + \varepsilon_{esc})\frac{l_s n_e}{A} \ . \tag{22}$$

We assume that the number density of the conductivity electrons is unchanged during the laser-matter interaction process. After insertion (17) into (22) the approximate formula for the ablation threshold takes the following form:

$$F_{th}^m \equiv I_0 t_p \approx \frac{3}{8}(\varepsilon_b + \varepsilon_{esc})\frac{c n_e}{\omega} \equiv \frac{3}{8}(\varepsilon_b + \varepsilon_{esc})\frac{\lambda n_e}{2\pi}. \tag{23}$$

The formula (23) predicts that the threshold fluence is proportional to the laser wavelength: $F_{th} \sim \lambda$. We demonstrate below that this relation agrees well with the experimental data.

### B. Ablation threshold for dielectrics

The ablation mechanism for the ionized dielectrics is similar to that for metals. However, there are several distinctive differences. First, an additional energy is needed to create the free carriers, i.e. to transfer the electron from the valence band to the conductivity band. Therefore, the energy equal to the ionization potential $J_I$, should be delivered to the valence electrons. Second, the number density of free electrons depends on the laser intensity and time during the interaction process as has been shown in Section IV. However, if the intensity during the pulse exceeds the ionization threshold then the first ionization is completed before the end of the pulse, and the number density of free electrons saturates at the level $n_e \sim n_a$, where $n_a$ is the number density of atoms in the target. Then the threshold fluence for ablation of dielectrics, taking into account the above corrections, is defined as the following:

$$F_{th}^d = \frac{3}{4}(\varepsilon_b + J_i)\frac{l_s n_e}{A} \tag{24}$$

Therefore, as a general rule, the ablation threshold for dielectric in the ultra short laser-matter interaction regime must be higher than that for the metals, assuming that all the atoms in the interaction zone are at least singly ionized. Because the absorption in the ionized dielectric also occurs in a skin layer, one can use the relation $l_s/A \approx \lambda/4\pi$ for the estimates and the scaling relations (see Appendix A).

Another feature of the defined above ablation thresholds (21) and (24) is that they do not depend explicitly on the pulse duration and intensity. However it is just a first order approximation. A certain, though weak, dependence is hidden in the absorption coefficient and in the number density of free electrons.

## 8. COMPARISON TO THE LONG PULSE REGIME

It is instructive to compare the above defined ablation threshold to that for the long laser pulses. This also helps in considering a general picture of the ablation process in a whole range of laser pulse duration.

The ultra short pulse laser-matter interaction mode corresponds to conditions when the electron-to-ion energy transfer time and the heat conduction time exceed significantly the pulse duration, $\tau_{ei} \sim t_{heat} \gg t_p$. Then the absorbed energy is going into the electron thermal energy, and the ions remain cold $\varepsilon_{ion} \ll \varepsilon_e$, making the conventional thermal expansion inhibited. However, as it was shown above, if the laser intensity is high enough, the electrons can gain the energy in excess of the Fermi energy and escape from the target. The electromagnetic field of the charge separation created by the escaped electrons pulls the ions out of the target. Hence, the extreme non-equilibrium regime of material ablation takes place. This regime occurs at the laser pulse duration $t_p < 200$ fs and at the intensities above $10^{13}$-$10^{14}$ W/cm$^2$. The escaped electrons accelerate the ions by the electrostatic field of charge separation.

An intermediate regime takes place at the laser pulse duration 0.5 ps $< t_p <$ 100 ps and at the intensities less than $10^{11}$ W/cm$^2$, when $\tau_{ei} \sim t_{heat} \sim t_p$, and $T_e \sim T_i$. The most appropriate description of the heating and expansion processes in this regime is given by the conventional two temperature approach [16].

At the longer laser pulse duration $t_p > 10$ ps the heat conduction and hydrodynamic motion dominate the ablation process, $t_p \gg \{\tau_{ei}; t_{heat}\}$. The electrons and the lattice (the ions) are in equilibrium early in the beginning of the laser pulse $T_e \sim T_i$. Hence, the limiting case of thermal expansion (thermal ablation) is suitable for the description of the long-pulse ablation mode. The ablation threshold for this case is defined by condition that the absorbed laser energy $AI_0 t_p$, is fully converted into the energy of broken bonds in a layer with the thickness of the heat diffusion depth $l_{heat} \sim (\kappa t_p)^{1/2}$ during the laser pulse [16]:

$$AI_0 t_p \cong (\kappa t_p)^{\frac{1}{2}} \varepsilon_b n_a. \tag{25}$$

The well-known $t_p^{1/2}$ time dependence for the ablation fluence immediately follows from this equation:

$$F_{th} \approx \frac{(\kappa t_p)^{1/2} \varepsilon_b n_a}{A}. \tag{26}$$

Equations (23), (24), and (26) represent two limits of the short- and the long-pulse laser ablation with a clear demonstration of the underlying physics. The difference in the ablation mechanisms for the thermal long pulse regime and the non-equilibrium short pulse mode is two-fold.

Firstly, the laser energy absorption mechanisms are different. The intensity for the long pulse interaction is in the range $10^8$-$10^9$ W/cm$^2$ with the pulse duration change from nanoseconds to picoseconds. The ionization is negligible, and the dielectrics are almost transparent up to UV-range. The absorption is weak, and it occurs due to the interband transitions, defects and excitations. At the opposite limit of the femtosecond laser-matter interaction the intensity is in excess of $10^{13}$ W/cm$^2$ and any dielectric is almost fully ionized in the interaction zone. Therefore, the absorption due to the inverse Bremmstrahlung and the resonance absorption mechanisms on free carriers dominates the interaction, and the absorption coefficient amounts to several tens percent.

Secondly, the electron-to-lattice energy exchange time in a long-pulse ablation mode is of several orders of magnitude shorter than the pulse duration. By this reason the electrons and ions are in equilibrium, and ablation has a conventional character of thermal expansion. By contrast, for the short pulse interaction the electron-to-ion energy exchange time, as well as the heat conduction time, is much larger than the pulse duration, and the ions remain cold. Electrons can gain energy from the laser field in excess of the Fermi energy, and escape the target. The electric field of a charge separation pulls ions out of the target thus creating an efficient non-equilibrium mechanism of ablation.

## 9. ABLATION DEPTH AND EVAPORATION RATE

The depth of a crater $x = d_{ev}$, drilled by the ultra short laser with the fluence near the ablation threshold $F = I_0 t > F_{th}$ (23) is of the order of the skin depth. According to Eq. (15), it increases logarithmically with the fluence:

$$d_{ev} = \frac{l_s}{2} \ln \frac{F}{F_{th}} \qquad (27)$$

due to the exponential decrease of the incident electric field and electron temperature in the target material. Equation (27) coincides apparently with that from [17]. However, one should note difference in definitions of the threshold fluence and skin depth in this paper from that in [17]. The skin depth calculated above for the laser interaction with copper target of 74 nm qualitatively complies with the ablation depth fitting to the experimental value of 80 nm [17].

The average evaporation rate, which is the number of particles evaporated per unit area per second, can be estimated for the ultra short pulse regime from (27) as the following:

$$(nv)_{short} = \frac{d_{ev} n_a}{t_p}. \qquad (28)$$

One can see a very weak logarithmic dependence on the laser intensity (or, fluence). For $d_{ev} \approx l_s \approx 70$ nm, $n_a \approx 10^{23}$ cm$^{-3}$, and $t_p \sim 100$ fs, one gets the characteristic evaporation rate for the short pulse regime of ~$7\times10^{30}$ $1/cm^2$ s. The evaporation rate for the long pulse regime depends only on the laser intensity [2]:

$$(nv)_{long} \approx \frac{I_a}{\varepsilon_b}. \qquad (29)$$

Taking $I_a \sim 10^9$ W/cm$^2$ and $\varepsilon_b \sim 4$ eV [2], the characteristic ablation rate for the long pulse regime of ~ $3\times10^{27}$ $1/cm^2$s is about $2\times10^3$ times lower.

The number of particles evaporated per short pulse $d_{ev}\times n_a \times S_{foc}$ ($S_{foc}$ is the focal spot area) is of several orders of magnitude lower than that for a long pulse. This effect eliminates the major problem in the pulsed laser deposition of the thin films, which is formation of droplets and particulates on the deposited film. The effect has been experimentally observed with 60 ps pulses and 76 MHz repetition rate by producing diamond-like carbon films with the rms surface roughness on the atomic level [2].

One also can introduce the number of particles evaporated per Joule of absorbed laser energy as a characteristic of ablation efficiency. One can easily estimate using Eqs.(28) and (29) that this characteristic is comparable for both the short-pulse and the long-pulse regimes.

## 10. COMPARISON TO THE EXPERIMENTAL DATA

Let us now to compare the above formulae to the different experimental data. Where it is available we present the full span of pulse durations from femtosecond to nanosecond range for ablation of metals and dielectrics.

### A. Metals

Let us apply Eq.(23) for calculation of the ablation threshold for Copper and Gold targets ablated by 780-nm laser. The Copper parameters are: density 8.96 g/cm$^3$, binding energy, e.g. heat of evaporation per atom $\varepsilon_b$ = 3.125 eV/atom, $\varepsilon_{esc}$ = 4.65 eV/atom, $n_a$ = $0.845\times10^{23}$ cm$^{-3}$. The calculated threshold $F_{th} \sim 0.51$ J/cm$^2$ is in agreement with the experimental figure 0.5-0.6 J/cm$^2$ [17], though the absorption coefficient was not specified in [17]. For the long pulse ablation taking into account thermal diffusivity of Copper 1.14 cm$^2$/s Eq. (26) predicts $F_{th}$ = 0.045[J/cm$^2$]×($t_p$ [ps])$^{1/2}$.

For a gold target ($\varepsilon_b$ = 3.37 eV/atom, $\varepsilon_{esc}$ = 5.1 eV, $n_e$ = 5.9×10$^{22}$ cm$^{-3}$) evaporated by laser wavelength 1053 nm the ablation threshold from Eq. (23) is $F_{th}$ = 0.5 J/cm$^2$. That figure should be compared to the experimental value of 0.45 ± 0.1 J/cm$^2$ [15]. For the long pulse ablation assuming the constant absorption coefficient of $A$ = 0.74 (see Appendix A) one finds from Eq.(26) $F_{th}$ = 0.049[J/cm$^2$]×($t_p$ [ps])$^{1/2}$. The experimental points [15] and the calculated curve are presented in Fig.1.

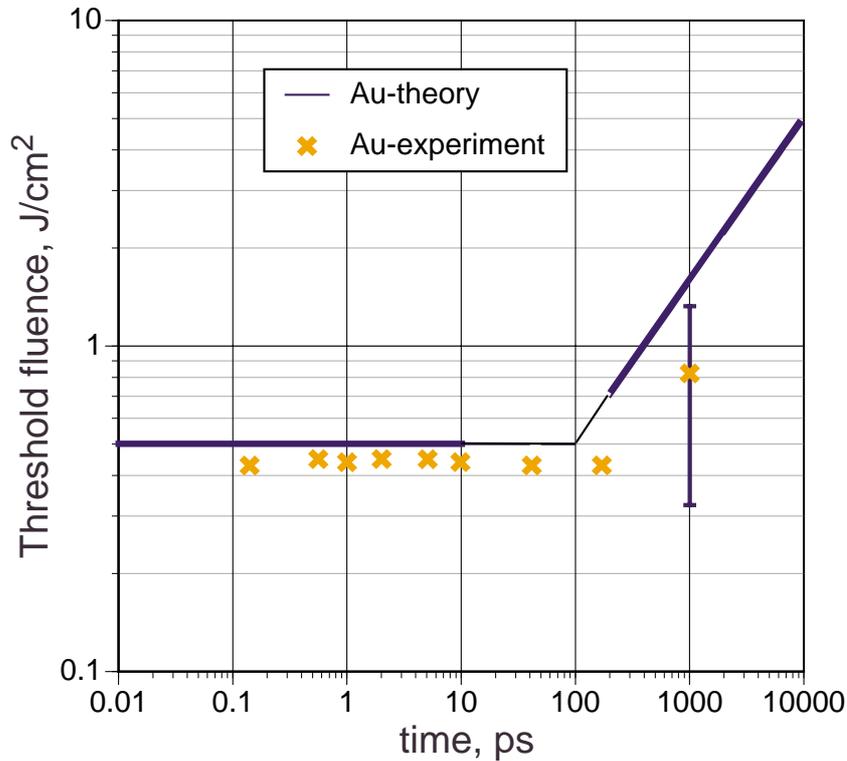

Fig.1. Threshold laser fluence for ablation of gold target versus laser pulse duration. The experimental error is ±0.5 J/cm$^2$ [15].

B. Silica

An estimate for the ablation threshold for silica from Eq.(26) ($n_e$ ~10$^{23}$ cm$^{-3}$, $\varepsilon_b$+$J_i$ ≈ 12 eV [24]) by a laser with $\lambda$ = 1053 nm ($\omega$ = 1.79×10$^{15}$ s$^{-1}$; $l_s/A$ ~ 83.8 nm) gives $F_{th}$ = 2.4 J/cm$^2$, which is in a qualitative agreement with the experimental figures ~2 J/cm$^2$ [1]. Formula (26) also predicts the correct wavelength dependence of the threshold: $F_{th}$ = 1.8 J/cm$^2$ for $\lambda$ = 800 nm ($l_s/A$ ~ 63.7 nm) and $F_{th}$ = 1.2 J/cm$^2$ for $\lambda$ = 526 nm (cf. Fig. 2). The experimental threshold fluences for the 100 fs laser pulse [1] are: 2 – 2.5 J/cm$^2$ ($\lambda$ = 1053 nm), ~ 2 J/cm$^2$ ($\lambda$ = 800 nm), and 1.2 – 1.5 J/cm$^2$ ($\lambda$ = 526 nm).

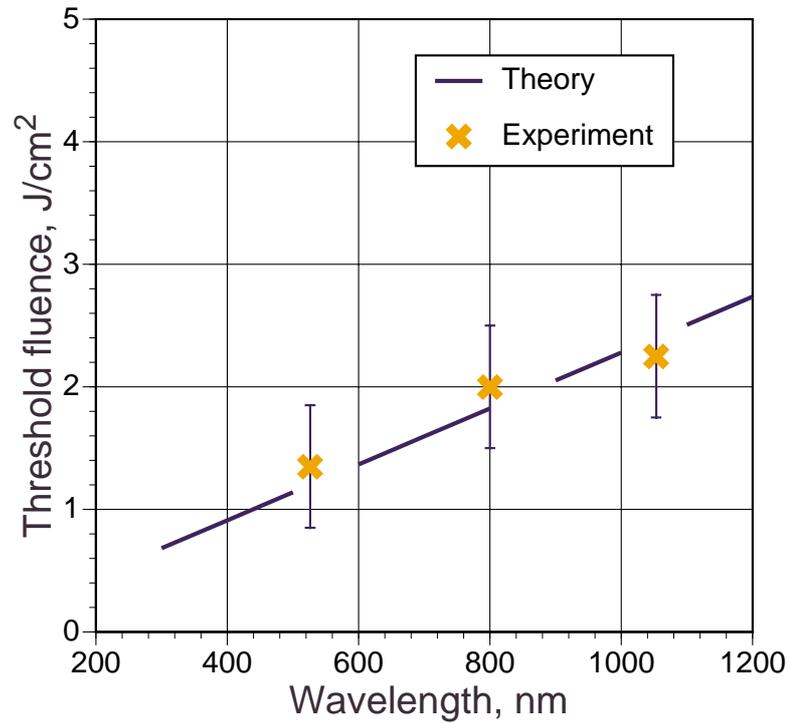

Fig.2. Threshold fluence for laser ablation of fused silica target as a function of the laser wavelength for 100 fs pulses. The experimental points are from the Ref. [1].

Using the following parameters for the fused silica at wavelength of 800 nm ($\kappa$ = 0.0087 cm$^2$/s, $\varepsilon_b$ = 3.7 eV/atom; $n_a$ = 0.7×10$^{23}$ cm$^{-3}$; and $A$ ~ 3×10$^{-3}$) one obtains a good agreement with the experimental data collected in [1] for the laser pulse duration from 10 ps to 1 ns. The long pulse regime Eq.(26) holds: $F_{th}$ = 1.29[J/cm$^2$]×($t_p$ [ps])$^{1/2}$ (see Fig. 3).

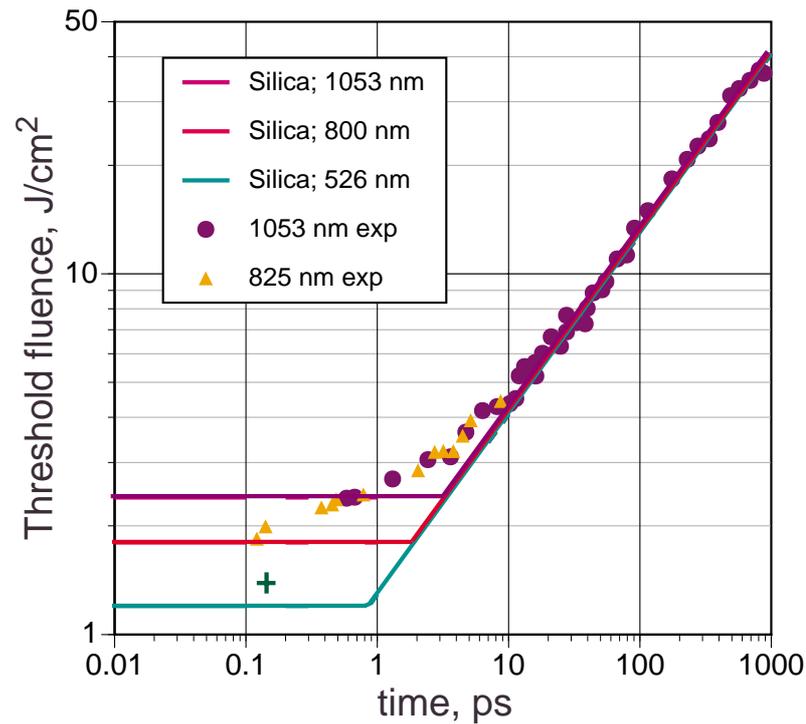

Fig. 3. Threshold laser fluence for ablation of fused silica target vs laser pulse duration. The experimental error is ±15% [1].

The ablation threshold of 4.9 J/cm$^2$ for a fused silica with the laser $t_p$ = 5 fs, λ = 780 nm, intensity ~10$^{15}$ W/cm$^2$ has been reported in [21]. This value is three times higher than that of [1] and from the prediction of Eq.(26). However, the method of the threshold observation, the absorption coefficient, as well as the pre-pulse to main pulse contrast ratio were not specified in [21].

In the Ref. [22] the crater depth of 120 nm was drilled in a BK7 glass by a 100-fs 620-nm laser at the intensity 1.5×10$^{14}$ W/cm$^2$. Assuming that the skin depth in the BK7 glass target is the same 84 nm as in the fused silica, the Eq.(29) for the ablation depth predicts the threshold value of 0.9 J/cm$^2$. This is in a reasonable agreement with the measured in [22] $F_{th}$ = 1.4 J/cm$^2$.

It should be noted that the definition of the ablation threshold implies that at the threshold condition at least a mono-atomic layer $x \ll l_s$, of the target material should be removed. Therefore, the most reliable experimental data for the ablation threshold are those obtained by the extrapolation of the experimental dependence of the ablated depth *vs* the laser fluence to the 'zero' depth. As one can see from above comparison, the experimental data on the ablation threshold determined this way are in excellent agreement with the formulae in this paper. It should be particularly emphasised that there were no any fitting coefficients in the calculations presented here.

## 11. DISCUSSION AND CONCLUSIONS

We described here a new regime of material ablation in the ultra short laser-matter interaction mode. The regime is characterised by the laser intensity in a range $\sim 10^{13} - 10^{14}$ W/cm$^2$ and the pulse duration shorter than the plasma expansion time, the heat conduction time, and the electron-to-ion energy transfer time. The interaction at such conditions results in ionisation of practically any target material. The interaction with the metals and dielectrics proceeds in a similar way in contrast to the conventional, long pulse interaction mode. The physics of this new regime of ablation consists in the ion acceleration in the electrostatic field created by hot electrons escaping from the target. We derived the explicit analytical formulae for the ablation threshold, the electron temperature in the skin layer, and the ablation rates for metals and dielectrics in terms of laser and target parameters. These formulae do not contain any fitting parameters and agree well with the available experimental data. In this new regime the threshold fluence is almost independent on the pulse duration, and the material evaporation rate is much higher than in the long pulse interaction regime.

An important condition for the ultra short pulse interaction mode in the real experiments is the high contrast ratio of the pulse: the target surface should not be ionized, damaged or ablated during the pre-pulse action. For the nanosecond-scale pre-pulse and the 100-fs main pulse the intensity contrast ratio must be of the order of $\sim 10^6$. The ultra short laser ablation can do a variety of fine jobs without any collateral damage to the rest of a target: cutting and drilling holes with a high precision, ablating all available materials with the ablation rate of several orders of magnitude faster than that with nanosecond lasers. The application of the ultra short lasers with high repetition rate for film deposition allows totally eliminate the problem of droplets and particulates on the deposited film. The theoretical background developed in this paper for laser ablation allows the appropriate laser parameters to be chosen for any given material and the laser-target interaction process to be optimized.

## APPENDIX A:

## ABSORPTION COEFFICIENT AND SKIN DEPTH NEAR THE ABLATION THRESHOLD

### 1. Metals: $\nu_{ei} \sim \omega_{pe} \gg \omega$

In these conditions the refraction coefficient expresses as the following:

$$N = n + ik \approx n(1+i); \quad n \approx k = \left(\frac{\omega_{pe}}{2\omega}\right)^{1/2}; \tag{A1}$$

the Fresnel absorption coefficient reads:

$$A = 1 - R \approx \frac{2}{n} - \frac{1}{n^2} = \left(\frac{8\omega}{\omega_{pe}}\right)^{1/2} \left\{1 - \left(\frac{\omega}{2\omega_{pe}}\right)^{1/2}\right\}; \tag{A2}$$

and correspondingly the skin-depth takes the form:

$$l_s = \frac{c}{\omega k} \approx c\left(\frac{2}{\omega \omega_{pe}}\right)^{1/2}. \tag{A3}$$

The ratio of $l_s/A$ that enters into the ablation threshold, expresses as the follows:

$$\frac{l_s}{A} \approx \frac{c}{2\omega}\left(1 - \left[\frac{\omega}{2\omega_{pe}}\right]^{1/2}\right)^{-1} = \frac{\lambda}{4\pi}\left(1 - \left[\frac{\omega}{2\omega_{pe}}\right]^{1/2}\right)^{-1}. \tag{A4}$$

Correction in the brackets for ablation of Copper ablation at 780 nm ($\omega = 2.415 \times 10^{15}$ s$^{-1}$; $\omega_{pe} = 1.64 \times 10^{16}$ s$^{-1}$) comprises 1.37. For a Gold target ablation at 1064 nm ($\omega = 1.79 \times 10^{15}$ s$^{-1}$; $\omega_{pe} = 1.876 \times 10^{16}$ s$^{-1}$) it amounts to 1.28. For the short wavelength such as KrF-laser or higher harmonics of Nd laser one should use the general formulae for the absorption coefficient and the skin-length.

### 2. Dielectrics: $\nu_{ei} \sim \omega_{pe} \sim \omega$

Repeating the above procedure for dielectrics one obtains $R \sim 0.05$, $A \sim 0.95$, and

$$\frac{l_s}{A} \approx \frac{3\lambda}{2\pi}. \tag{A5}$$

# APPENDIX B: IONIZATION OF SILICA

The ionization potential of Si is $J_i = 8.15$ eV. For Nd:YAG laser ($\lambda = 1064$ nm) at the intensity $2 \times 10^{13}$ W/cm$^2$ the probability for the ionization by electron impact is $w_{imp} = 10^{13}$ s$^{-1}$, for the multiphoton ionization is $w_{mpi} = 5 \times 10^{-4}$ s$^{-1}$, and the number density of created free electrons in 100 fs is $n_e \sim 10^7$. At the intensity $10^{14}$ W/cm$^2$ $w_{imp} = 10^{13}$ s$^{-1}$; $w_{mpi} = 5 \times 10^{14}$ s$^{-1}$; and the number density of free electrons reaches the solid density $n_e \sim n_a \sim 10^{23}$ cm$^{-3}$ in 20 fs – this is the time required for full first ionization.